\documentclass[aps,twocolumn,prb,showpacs,preprintnumbers,amsmath,amssymb,superscriptaddress]{revtex4}
\usepackage{mathptm}  
\usepackage{dcolumn}                    
\usepackage{bm}                        
\usepackage{graphicx}
\usepackage{times}
\begin{document}
\newcommand {\be}{\begin{equation}}
\newcommand {\ee}{\end{equation}}
\newcommand {\bea}{\begin{eqnarray}}
\newcommand {\eea}{\end{eqnarray}}
\newcommand {\nn}{\nonumber}
\newcommand {\bb}{\bibitem}
\small

\title{Theory of tunneling spectroscopy in UPd$_{2}$Al$_{3}$}

\author{David Parker}
\affiliation{Max Planck Institute for the Physics of Complex Systems,
N\"{o}thnitzer Str. 38, D-01187 Dresden, Germany}

\author{Peter Thalmeier}

\affiliation{Max Planck Institute for the Chemical Physics of Solids,
N\"othnitzer Str. 40, D-01187 Dresden, Germany}

\date{\today}

\begin{abstract}

There is still significant debate about the symmetry of the order parameter in
the heavy-fermion superconductor UPd$_{2}$Al$_{3}$, with proposals for $\cos(k_{3})$,$\cos(2k_{3})$, $\sin(k_{3})$, and $e^{i\phi}\sin(k_{3})$.  Here we analyze the tunneling spectroscopy of this compound and demonstrate that the experimental results by Jourdan et al \cite{jourdan} are inconsistent with the
last two order parameters, which are expected to show zero-bias conductance
peaks.  We propose a definitive tunneling experiment to distinguish between the first two order parameters.
\end{abstract}
\maketitle
\section{Introduction}

Superconductivity in UPd$_{2}$Al$_{3}$ with a transition temperature T$_{c}$
= 2 K was discovered in 1991 by Geibel et al \cite{geibel}, and since that
time much experimental and theoretical work has been performed, with a 
particular aim of establishing the order parameter symmetry.  
In general, this determination is a crucial first step in understanding 
the pairing mechanism of a given superconductor, so that large efforts
are generally expended in this direction.

Evidence for unconventional or nodal superconductivity in UPd$_{2}$Al$_{3}$
has emerged from a variety of experiments.  Feyerherm et al \cite{feyerherm}
measured the Knight shift in UPd$_{2}$Al$_{3}$ and found a substantial 
reduction below T$_{c}$, indicative of singlet pairing.
In NMR experiments Tou \cite{tou} and Matsuda \cite{matsuda} 
found low-temperature
nuclear spin lattice relaxation rate T$_{1}^{-1}$ power law (T$^{3}$) behavior, and the 
absence of a Hebel-Slichter coherence peak below T$_{c}$.  Both behaviors 
are characteristic of nodal superconductivity, and in particular the 
observation of T$^{3}$ behavior 
is strongly suggestive of line nodes in the order parameter.  
Similar evidence was obtained by Sato \cite{sato}, who measured the 
low-temperature specific heat and found T$^{2}$ behavior, also indicative
of line nodes.  Hiroi et al \cite{hiroi} measured the low-temperature
thermal conductivity and also found T$^{2}$ behavior.  More recently,
Watanabe et al \cite{watanabe}  measured the
angle-dependent magnetothermal conductivity and found evidence for
a single line node parallel to the basal plane of the hexagonal
UPd$_{2}$Al$_{3}$.

Tunnelling spectroscopy can be a powerful probe of order parameter symmetry.  
The pioneering work in this field was performed by Blonder, Tinkham and 
Klapwjik (BTK) \cite{blonder}, who provided a simple solution to the problem of Andreev reflection \cite{andreev} in s-wave superconductors in an N-I-S contact, and by Tanaka et al
\cite{tanaka}, who was able to explain the zero-bias conductance peaks (ZBCP)
generally observed in tunneling measurements of the high-temperature cuprate 
superconductors \cite{wei-yeh}.  Since that time much theoretical work on tunneling spectroscopy in various materials has been performed; primary references of interest are the book by Tinkham \cite{tinkham} and the study by Honerkamp and Sigrist \cite{hongerkamp}.

The basic conclusion of Tanaka's original work was that zero-bias conductance
peaks are intimately tied in with the phase difference between the 
pair potentials of the transmitted electron-like and hole-like quasiparticles
on the superconducting side of the junction.  When this phase difference
reaches $\pi$, Andreev reflection (resulting in the transmission 
of two electrons for a single incoming electron) is enhanced
and normal reflection diminished, so that the transmission coefficient is
increased.  This enhancement is most prominent at zero-energy.  At increasing tunneling energy barrier height (the thin oxide 
layer between the superconductor and normal) the normal state conductance
decreases, so that the relative conductance, or ratio of superconducting to normal state conductance, increases.  It should be noted that while strongly
enhanced ZBCPs (diverging in the limit of large barrier height) only 
result from a $\pi$ phase difference, significant zero-energy states can be
realized in other cases, as in \cite{hongerkamp}.  We show below that for the case of interest, substantial amounts of zero-energy states can be realized even for order parameters
without the requisite $\pi$ phase change, when the effects of quasiparticle
lifetime are accounted for.  The magnitude of these zero-energy states, as well
as the height and energy of the quasiparticle coherence peaks, are strong
functions of the order parameter, and for the geometry appropriate to UPd$_{2}$Al$_{3}$, of the normal metal used in the junction, so that a wide range of interesting and 
informative experimental results can be obtained, offering a definitive resolution to the question of order parameter symmetry in UPd$_{2}$Al$_{3}$.

\section{Model}

In general, the pair state of a superconductor is described by the
Bogoliubov-deGennes equations \cite{deGennes_book, blonder,tanaka,klapwijk_mar}:
\bea
i\hbar\frac{\partial f}{\partial t} &=& -\left[\frac{\hbar^{2}\nabla^{2}}{2m}+
\mu+V(x)\right]
f({\bf x},{\bf k},t) - \Delta({\bf x},{\bf k})g(x,t) \\
i\hbar\frac{\partial g}{\partial t} &=& \left[\frac{\hbar^{2}\nabla^{2}}{2m}+\mu+V(x)\right]g({\bf x},{\bf k},t) 
- \Delta({\bf x},{\bf k})f(x,t)
\eea
Here, as in Klapwijk et al \cite{klapwijk_mar}, we take the f's as electron
wave functions and the g's as hole wave functions, and note that the solutions
to these equations can be written as
\bea
f({\bf x},{\bf k},t)&=& u({\bf k})\exp(i\ ({\bf k}\cdot {\bf r} - Et)/\hbar) \\
g({\bf x},{\bf k},t)&=& v({\bf k})\exp(i\ ({\bf k}\cdot {\bf r} +Et)/\hbar)
\eea
where u and v are the well-known BCS coherence factors \cite{BCSbook,
hongerkamp, klapwijk_mar}:
\bea
u({\bf k}) &=& \sqrt{\frac{1}{2}(1+\sqrt{E^{2}-|\Delta^{2}({\bf k})|}/E)} \\
v({\bf k}) &=& \sqrt{\frac{1}{2}(1-\sqrt{E^{2}-|\Delta^{2}({\bf k})|}/E)}
\eea
We note that in the initial BCS formulation these u and v coefficients are
only defined for $|E| > |\Delta({\bf k})|$, 
but in this work we will extend this
usage to energies $E < \Delta({\bf k})$ including negative energy, and will 
model the effects of finite quasiparticle lifetime by letting $ E \rightarrow
E - i\Gamma$ \cite{dynes}, where $\Gamma$ is the quasiparticle scattering rate.
Note that this produces complex coherence factors whose squared amplitudes 
do not, in
general, sum to unity.  This is a consequence of the fact that the
transmitted waves in the superconductor are evanescent waves for 
$E < \Delta({\bf k})$ and decay rapidly, with the electrons transmitted
to the condensate \cite{blonder}.  These waves are effectively surface states.
This extension of the coherence factors is crucial for studying the
effects of tunneling voltages $V < \Delta({\bf k})$, as are measured.  

With the coherence factors in place, we are now in a position to consider the
process whereby an incident electron in a normal metal undergoes Andreev and
normal reflection at the N-S interface, with an electron-like quasiparticle
and a back-scattered hole-like quasiparticle transmitted.  This process is
depicted in Figure 1.  Note that the arrows indicate the particle group 
\begin{figure}[h!]
\includegraphics[width=7.5cm]{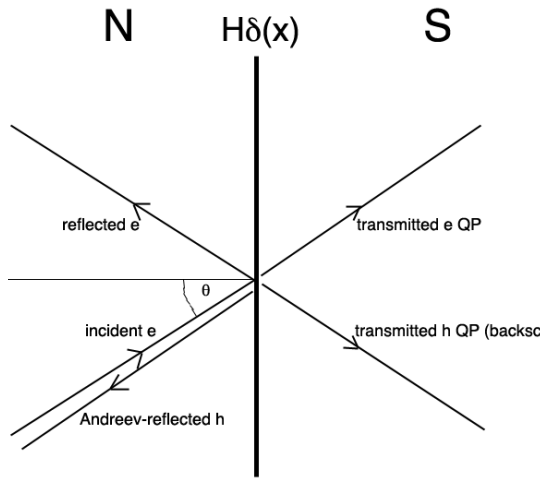}
\caption{A schematic diagram of all the particles involved in the
reflection/transmission process at a N-S boundary; the arrows indicate
the direction of the group velocity.  Note that for the holes the
momenta are opposite to this direction.}
\end{figure}
velocity; for the transmitted hole and Andreev-reflected hole the momentum is
opposite to this.  We have included a delta-function potential
$H\delta({\bf x})$ to model the inclusion of a thin oxide layer
between the superconducting UPd$_{2}$Al$_{3}$ and the normal metal.  
Translational invariance parallel to the interface dictates that the momentum
in this direction be conserved, while the momentum perpendicular to the
interface is also conserved in the approximation where the barrier energy is
much less than the Fermi energy of the incoming electron.

We note that unlike in virtually all other studies of the tunneling spectroscopy of
unconventional superconductors, the Fermi momentum of the normal metal
plays a key role in selecting the wavevectors which will contribute
to the conductance, as indicated in Figure 2. Only a very small portion
\begin{figure}[h!]
\includegraphics[width=8.5cm]{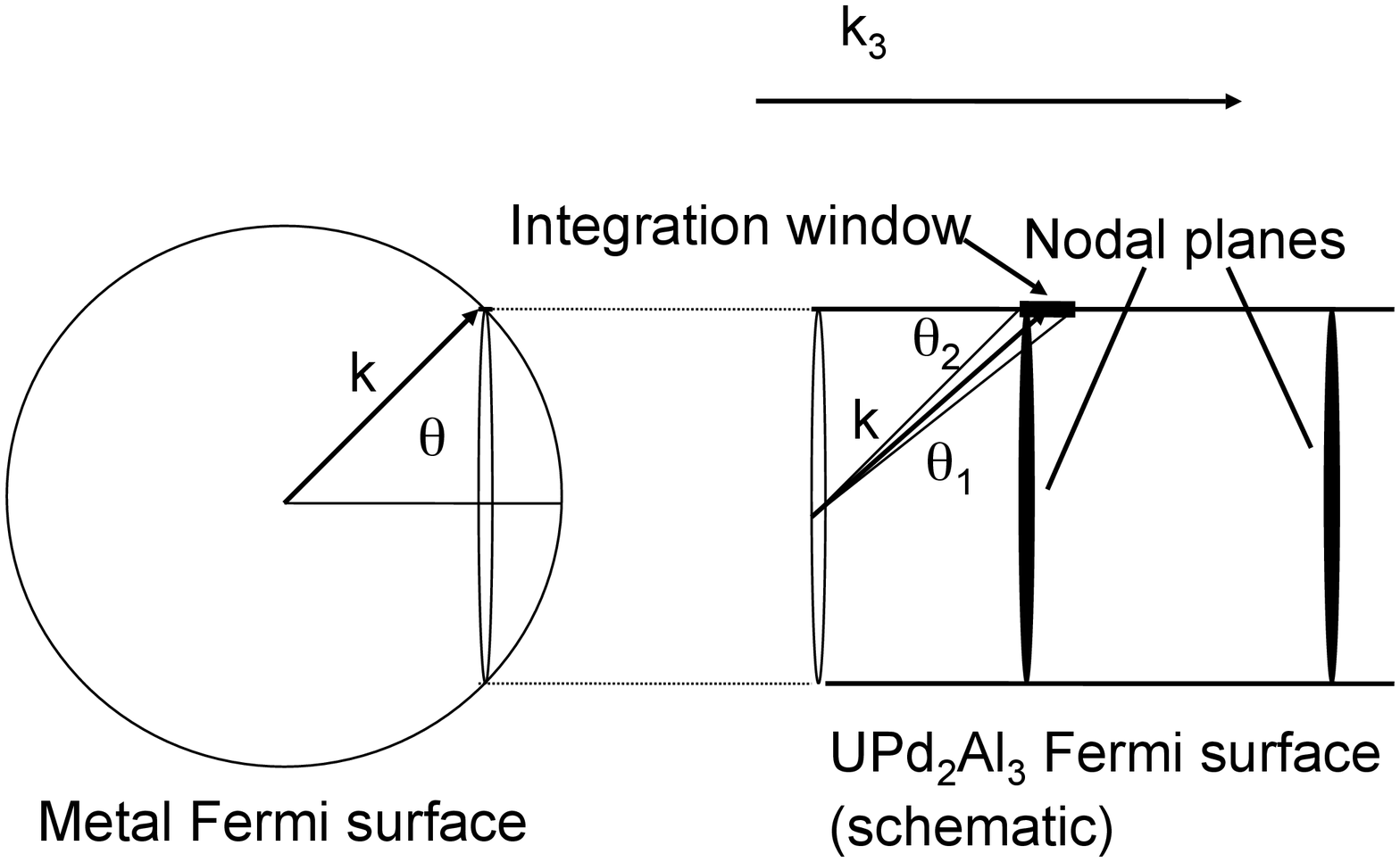}
\includegraphics[width=8.5cm]{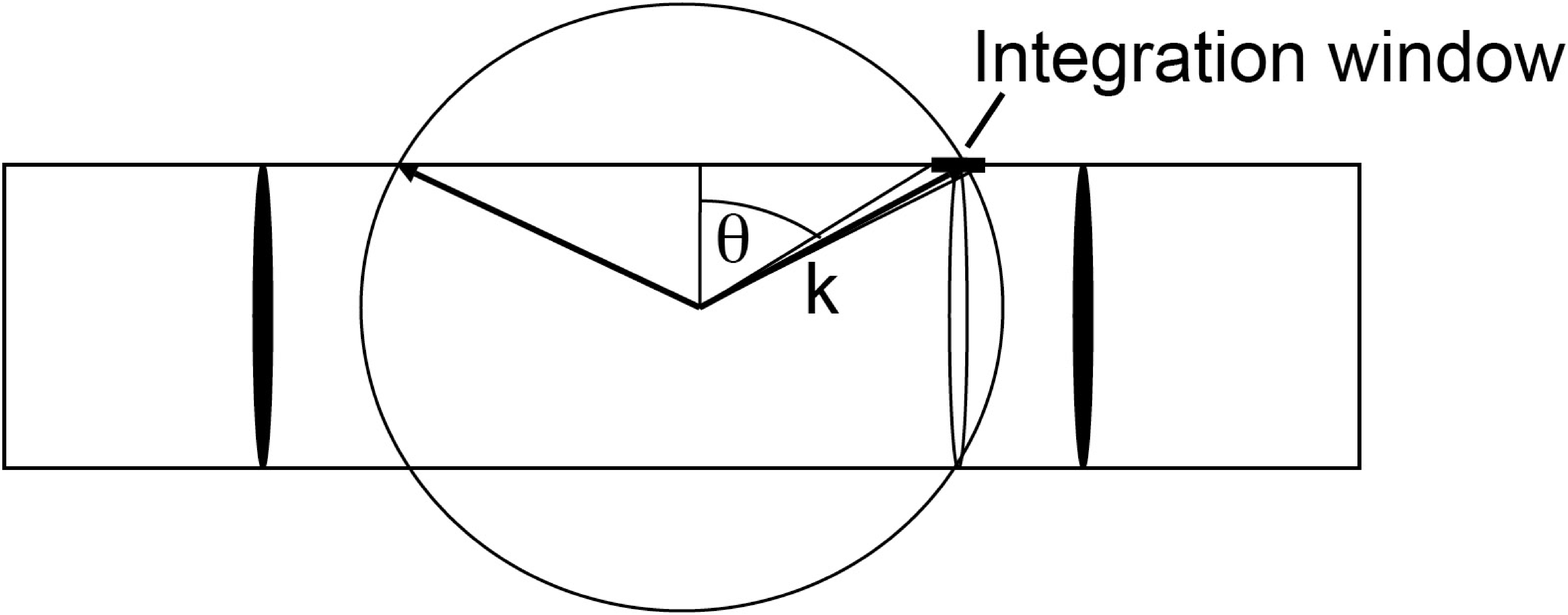}
\caption{A schematic diagram of the metallic and UPd$_{2}$Al$_{3}$ Fermi
surfaces, for the longitudinal and transverse geometries.  
For the UPd$_{2}$Al$_{3}$ Fermi
surface we take the main cylindrical sheet 
\cite{deHaas papers} and neglect its corrugation.}
\end{figure}
of the electrons in a spherical Fermi surface approximation will lie on the 
dominating, approximately cylindrical Fermi surface of UPd$_{2}$Al$_{3}$, as indicated, and
therefore contribute to the superconducting or normal state conductance.
Electrons far from the Fermi surface cylinder of the UPd$_{2}$Al$_{3}$ will play no 
role.

We will also see that the conductance results depend critically on whether
the wavevector selected lies near a nodal line; thus different metals 
used on the normal side can be expected to display radically different behavior.  Indeed, we will see that the measured gap magnitude, the zero-energy
density-of-states, and even the existence of the usual coherence peaks will
depend strongly upon the metal used.  Finally, we have corrected for the effects of the electron effective mass mismatch \cite{yakovenko} at the interface.  

\section{Calculation methodology}

In our calculations we follow the work of Blonder et al \cite{blonder}, Honerkamp and Sigrist \cite{hongerkamp}, and Tanaka \cite{tanaka}.  The basic
method is straightforward: apply the usual boundary conditions at the interface
and solve for the ampltitudes of the various components. 

The incoming electron wavefunction can be written as $\binom{1}{0}
e^{ik_{F}x\cos\theta}$. with $\theta$ the angle between the normal to the 
interface and ${\bf k}$.  We have suppressed the part of the wavefunction
depending on the momentum parallel to the interface, as due to the 
translational invariance in this direction all of the involved momenta
are equal.  On the normal metal side, the Andreev-reflected hole, with 
amplitude $a$, has wavefunction $\binom{0}{1}e^{ik_{F}x\cos\theta}$, 
and the normally-reflected electron, with amplitude $b$, has wavefunction
$\binom{1}{0}e^{-ik_{F}x\cos\theta}$.  On the superconducting side, the
electron-like quasiparticle, with amplitude $c$, has wavefunction 
$\binom{u(\theta)}{\exp(-i\phi_{\theta})v(\theta)}e^{ik_{F}x\cos\theta}$, 
and the backscattered hole-like quasiparticle has wavefunction 
$\binom{\exp(i\phi_{\pi-\theta})v({\pi-\theta})}{u({\pi-\theta})}
e^{ik_{F}x\cos\theta}.$  Here the exponential factors represent
the phase of the gap at the indicated angle.  We are now able to apply the boundary conditions.
Continuity of the wavefunction at the interface yields
the first two of four equations for the four unknowns a,b,c and d. 

The usual boundary condition \cite{blonder,tanaka} appropriate for a delta-function potential $H\delta(x)$ is
\bea
\psi_{S}^{'}(0)-\psi_{N}^{'}(0) &=& \frac{2m}{\hbar^{2}}H\psi(0)
\eea
However, due to the large effective mass mismatch at the boundary ($m_{S}
\sim 100 m_{0}$) we must generalize this condition.  Integrating 
the Bogolubov-deGennes equations across the boundary we find
\bea 
\frac{\psi_{S}^{'}(0)}{m_{S}}-\frac{\psi_{N}^{'}(0)}{m_{N}}
 &=& \frac{2}{\hbar^{2}}H\psi(0)
\eea
with the effective masses as indicated.  Substituting
the wavefunctions outlined
above into this formula leads to the final two equations
for the four unknowns a, b, c, and d.  We find the following solutions for $a$
and $b$:
\bea 
a(\theta,E)=\frac{uv}{D}
\eea
with
\bea
D = (u^{2}-e^{i(\phi_{\pi-\theta}-\phi_{\theta})}v^{2})|Z|^{2}m+ u^{2}\\
b(\theta,E) =\frac{(u^{2}-e^{i(\phi_{\pi-\theta}-\phi_{\theta})}v^{2})(|Z|^2+Z)}{D}
\eea
Here $m=m_{S}/m_{N}$ and $Z=H/(i v_{F}\cos\theta)+1/2m-1/2$ where $v_{F}$ is the metallic
Fermi velocity, and $e^{i\phi}=\frac{\Delta(\phi)}{|\Delta(\phi)|}$, i.e. the
gap phase.  The energy dependence of a and b is contained in the coherence factors u and v.  We note two interesting effects: 

\begin{itemize} 
\item The denominator D increases rapidly with mass mismatch m and
effective barrier height Z.  There are therefore narrow transmission 
resonances when the prefactor of the 
$|Z|^2 m$ term vanishes.  For all the order parameters under
consideration, the factors $u_{\theta}$ and $u_{\pi-\theta}$ are identical,
as are $v_{\theta}$ and $v_{\pi-\theta}$.  When the phase factor is unity,
i.e. there is no change of order parameter sign under the change in angle
from $\theta$ to $\pi-\theta$, the prefactor can only vanish when u=v, i.e. $E=\Delta({\bf k})$.  Thus for the two cosine-containing order parameters, we expectcoherence peaks
in the differential conductance at the gap energy 
selected by the wavevector matching on the UPd$_{2}$Al$_{3}$ Fermi surface.  
This implies
that the energy, or voltage, of the conductance coherence peaks for these order parameters is a function of the normal metal's Fermi 
wavevector, in contrast to most tunneling spectroscopy studies, where no effects of wavevector matching are considered, because the Fermi surface on
the superconducting side is also considered spherical.  It is the peculiar geometry engendered
by the longitudinal Fermi surface of UPd$_{2}$Al$_{3}$ that creates 
this unusual coherence peak effect.
\item For the two sine-containing order parameters, the phase factor
is -1, so that for a resonance we must satisfy $u^{2}=-v^{2}$.  Strictly
speaking this is impossible, since $u^{2}+v^{2}=1$, but the factors u and v
diverge as $|E| \rightarrow 0$, so that both the numerator of $a(E,\theta)$
and the second term of D diverge, while the first term of D remains finite.
As $E \rightarrow 0$, explicit calculation shows that, for $\Gamma=0$ 
we attain perfect transmission, and
therefore, we expect zero-energy resonances to appear as zero-bias
conductance peaks for these two order parameters.
\end{itemize}
Following BTK and Tanaka \cite{blonder,tanaka}, the normalized conductance at temperature
$T=1/\beta$ is given by 
\bea
dI/dV = \sigma(V) \equiv \frac{\sigma_{S}(V)}{\sigma_{N}(V)}
\eea
\bea
= \frac{\int_{-\infty}^{\infty}dE\int_{\theta_{1}}^{\theta_{2}}d\theta
\mathrm{sech}^2(\beta(E-V)/2)(1+|a(\theta,E)|^{2}-|b(\theta,E)|^{2})}{\int_{-\infty}^{\infty}dE\int_{\theta_{1}}^{\theta_{2}}d\theta
\mathrm{sech}^2(\beta(E-V)/2)\sigma_{N}(E))}
\eea
Note that we are not integrating over the full range of angles $\theta$ as the majority of such angles will result in
wavevectors lying away from the nearly cylindrical \cite{deHaas papers} Fermi surface of UPd$_{2}$Al$_{3}$.  In practice
what we have done in all cases is to use the metallic Fermi momentum to determine the angle at which this momentum
lies on the UPd$_{2}$Al$_{3}$ Fermi surface and then 
computed $k_{z0}$ (the component of this momentum along the Fermi surface).  As a 
rough accounting for actual metallic Fermi surface deviations from spherical we have then integrated over angles
corresponding to $k_{z}$ from 0.95 to 1.05 $k_{z0}$.  For alkali metals 
such as sodium, \cite{ashcroft} this is an 
overestimation, while for other materials - such as 
tungsten - this understates the Fermi surface deviations from spherical.  
Nevertheless, the results of our calculations are not sensitive to the 
size of the integration window used, provided that this window is small compared with the distance between the nodal planes.  For all calculations we have taken the effective mass mismatch m as 100, while the barrier height $H/v_{F}$
has been taken as unity.  The results are not sensitive to either of these
parameters.

\section{Order parameter proposals}

Before presenting the results of our calculations we give a brief summary
of the order parameters which have been proposed for UPd$_{2}$Al$_{3}$.  Note
that in all proposals $k_{3}$ is taken as perpendicular to the ab-plane, and c
is the c-axis lattice constant.

Won et al \cite{parker} proposed $\Delta({\bf k})=\cos(2ck_{3})$ based upon
magnetothermal conductivity data \cite{watanabe}, while McHale et al \cite{mchale} performed a strong-coupling calculation and found that both
$\cos{ck_{3}}$ and $\sin{ck_{3}}$ had the highest transition temperatures,
among possible order parameters.  It was noted \cite{ymatsuda} that
the chiral order parameter $\exp(i\phi)\sin{ck_{3}}$
where $\exp(i\phi)=\hat{k_{1}}+i\hat{k_{2}}$, is also a possible candidate.
All of these order parameters contain nodal lines perpendicular to the 
$k_{3}$ axis, which were found in magnetothermal conductivity \cite{watanabe}
measurements on this material.

We note that the last two order parameters are odd under reflection around
the basal plane; for all order parameters
this reflection is equivalent to replacing $\theta$ in Figs.
1 and 2 and the above equations by $\pi-\theta$ (note that this is equivalent
to reflection around $\theta=\pi/2$).  This is immediately apparent for 
$\sin{ck_{3}}$, and the geometric diagram below shows that it is true for 
$\exp(i\phi)\sin{ck_{3}}$.  Hence,
\begin{figure}[h!]
\includegraphics[width=8cm]{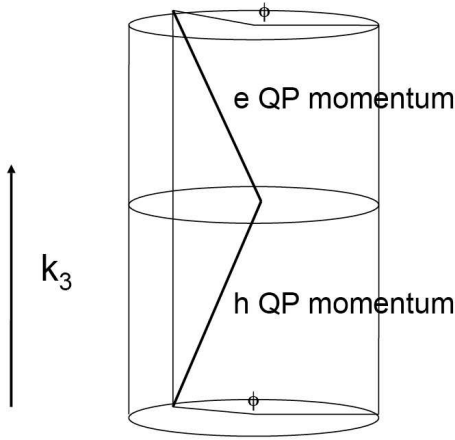}
\caption{A schematic diagram of the wavevector geometry for
the chiral order parameter
$\Delta({\bf k}) = e^{i\phi}\sin(ck_{3})$. Here k$_{3}$ is along the vertical axis.  Note that the transmitted hole-like
and electron-like quasiparticles have the same value of the phase $\phi$, so
that no additional phase is introduced by the chiral factor and the results
are identical to the $\sin(ck_{3})$ case.}
\end{figure}
when the tunneling is along the $k_{3}$ axis, this sign change is expected
to produce zero-bias conductance peaks (ZBCP) \cite{tanaka} in the tunneling conductance.  We will see that this is indeed the case.

\section{Results}

Below in Fig. 4 
is presented the normalized differential conductance $dI/dV$ for the
four order parameters described above, for the case where the tunneling
direction is along the c-axis and the normal metal is lead, and $T \ll T_{c}$.
These are the conditions of the experimental data at T = 0.3K by Jourdan et al \cite{jourdan}.  We have here taken the quasiparticle lifetime
broadening factor $\Gamma$ to be 54 $\mu V$, which compares
well with that extracted from de Haas-van Alphen measurements 
\cite{deHaas papers}, which measured Dingle temperatures of 0.1 - 0.28 K,
corresponding to $\Gamma = 27-76 \mu V$. We find excellent agreement with the data of Jourdan et 
al \cite{jourdan} throughout the whole range modeled for the order
parameters $\cos(ck_{3})$ and $\cos(2ck_{3})$, while contrary to experiment, a
ZBCP is predicted for the two sine-containing order parameters.  
\begin{figure}[h!]
\includegraphics[width=8cm]{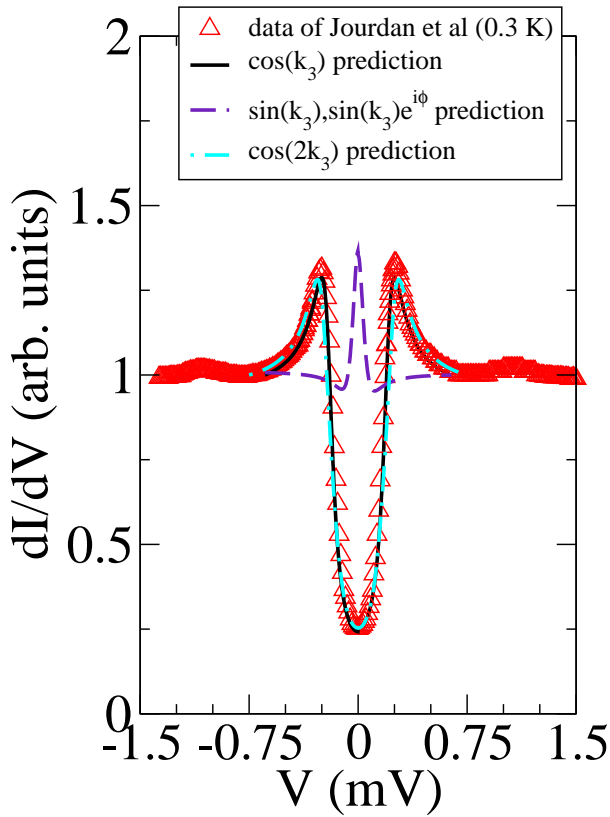}
\caption{The tunneling data of Jourdan at 0.3 K, with lead as the normal metal, is compared with predictions
for the four order parameters shown.}
\end{figure}
Based on this data
we thus believe it is rather unlikely that either of these two order parameters
is relevant for UPd$_{2}$Al$_{3}$, although based upon calculations by 
Nishikawa \cite{nishikawa} the $\sin(ck_{3})$ gap function may have
relevance for the isostructural superconductor UNi$_{2}$Al$_{3}$.  This material
has indeed been proposed as a triplet superconductor having an 
odd gap function. \cite{ishida}

For completeness, we have computed dI/dV in the transverse direction,
where the tunneling current is in the basal plane, with the same parameters
as above.  The results are shown in Fig. 5.  As in the first plot, there is little difference between the two
cosine order parameters, while the $\sin(ck_{3})$ result shows 
virtually no coherence peak and no ZBCP, but a high zero-energy density-
of-states (ZDOS).  This is a direct result of the selected angle of incidence
for lead in this geometry falling very near the nodal line for this order
parameter; this property will be used to full advantage in proposing an
experiment to distinguish between the two cosine order parameters.
\begin{figure}[h!]
\includegraphics[width=7.4cm]{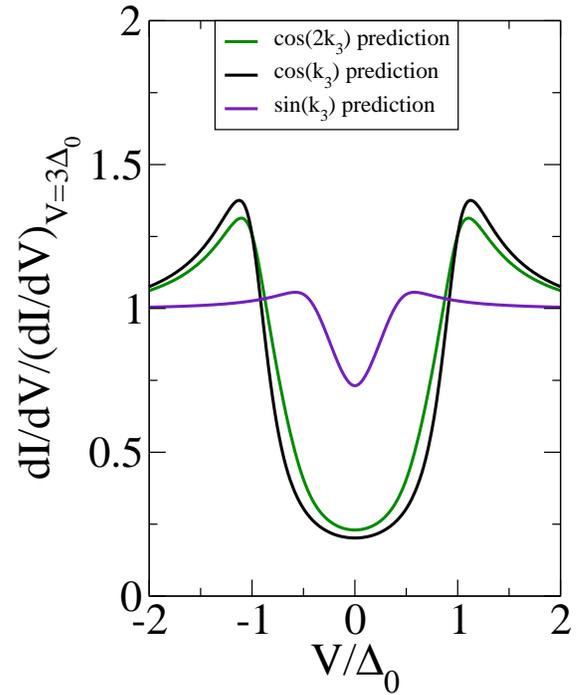}
\caption{The predictions for the ab-plane tunnelling conductance for the three
order parameters indicated are shown.}
\end{figure}
We have also performed calculations for the $\cos(ck_{3})$ gap function for the entire range of temperatures tested
by Jourdan et al, from 0.3 K to 1.5 K (in this experiment, T$_{c}$ was 1.6 K.).
Figure 6 below shows excellent agreement with this order parameter over
the entire temperature range, for nearly all energies  We stress that this was accomplished with a 
minimum of fitting parameters; the quasiparticle lifetime broadening rate
$\Gamma$ was assumed temperature-independent (=54$\mu$V) and the maximum 
energy gap value $\Delta(T)$ for all temperatures was near the value
\begin{figure}[h!]
\includegraphics[width=7.4cm]{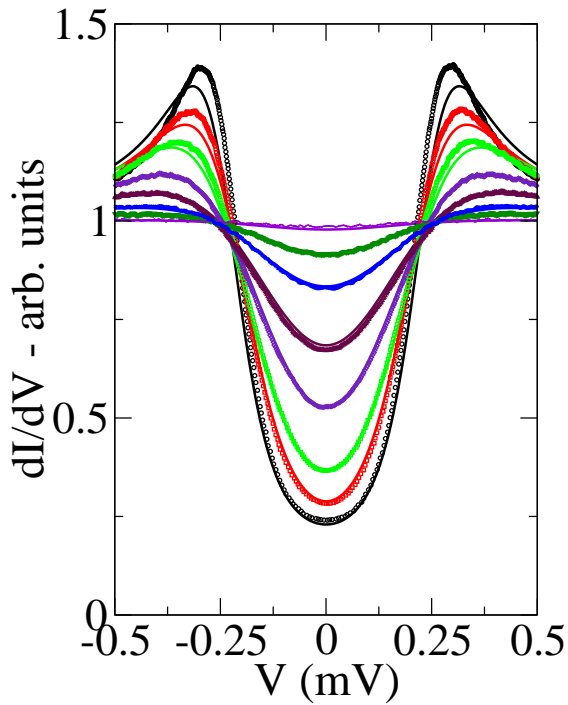}
\caption{The fits to the data of Jourdan using the $\cos(ck_{3})$ order parameter are shown.}
\end{figure}
appropriate for BCS weak-coupling superconductors of this pairing symmetry \cite{footnote,won}.
Below we have plotted the fitted values of $\Delta(T)$ compared
\begin{figure}[h!]
\includegraphics[width=8cm]{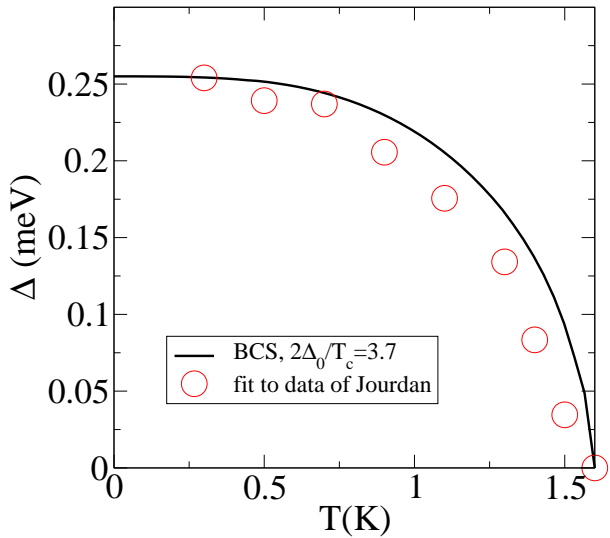}
\caption{The fitted values of $\Delta(T)$ are shown.}
\end{figure}
to the BCS prediction.  .
We note that based upon the fits of these data,
the ratio $2\Delta(0)/T_{c}$ is approximately 3.7, indicating that
a weak-coupling treatment may be accurate for this calculation.  For
this order parameter, in the weak-coupling limit the ratio $2\Delta(0)/T_{c}$
is 4.3.  We also note,
however, the slight bump in the data of Jourdan in Fig. 4 around $V=1 mV$, which
may be indicative of strong-coupling to a magnetic-exciton boson 
\cite{mchale,thal_latest}. 

We have not performed detailed comparison with experiment for the $\cos(2ck_{3})$ order parameter, although it is quite possible that similar agreement with
experiment could be obtained for this case.  We believe that this order
parameter may be less likely to apply for UPd$_{2}$Al$_{3}$ for the reasons
presented in \cite{thal_latest}; namely, the formation of a resonance peak
in inelastic neutron scattering \cite{ins} requires, in the magnetic exciton
scenario, a sign change over points on the Fermi surface separated
by the antiferromagnetic
wave-vector ${\bf q}=\frac{\pi}{c}\hat{k_{3}}$ and this does not apply
for this order parameter.  Nevertheless, we have devised what we believe is a
definitive experiment to finally determine the order parameter 
in UPd$_{2}$Al$_{3}$ after fifteen years \cite{geibel} of study.  

\section{Proposed Experiment - Order Parameter Determination}

The proposed experiment makes use of the unusual Fermi surface geometry
effects present for tunnelling in the c-axis direction.  The matching of 
wavevectors in the metal and in UPd$_{2}$Al$_{3}$ places sharp constraints on 
which region of the UPd$_{2}$Al$_{3}$ Fermi surface is employed by the
superconducting electrons.  In particular, it is possible to choose the normal
metal so that for the $\cos(2ck_{3})$ order parameter, 
the selected wavevector lies very near the nodal line, while for the
$\cos(ck_{3})$ order parameter, the selected wavevector is much more distant
from the nearest node.  We have chosen Calcium for this purpose as its
Fermi wavevector, based on a spherical Fermi surface approximation 
\cite{ashcroft}
lies very near a nodal line of $\cos(2ck_{3})$ but is far from a nodal
line of $\cos(ck_{3})$. 
The same situation would apply for Thallium and Cesium, but these are
soft and reactive metals \cite{CRC} 
that are not likely to be useful in forming a tunnel junction.

In Figure 8 we show the results of the calculation for calcium at T=0, with the
same quasiparticle lifetime broadening factor as before.  We observe
that for the $\cos(ck_{3})$ case, 
\begin{figure}[h!]
\includegraphics[width=8cm]{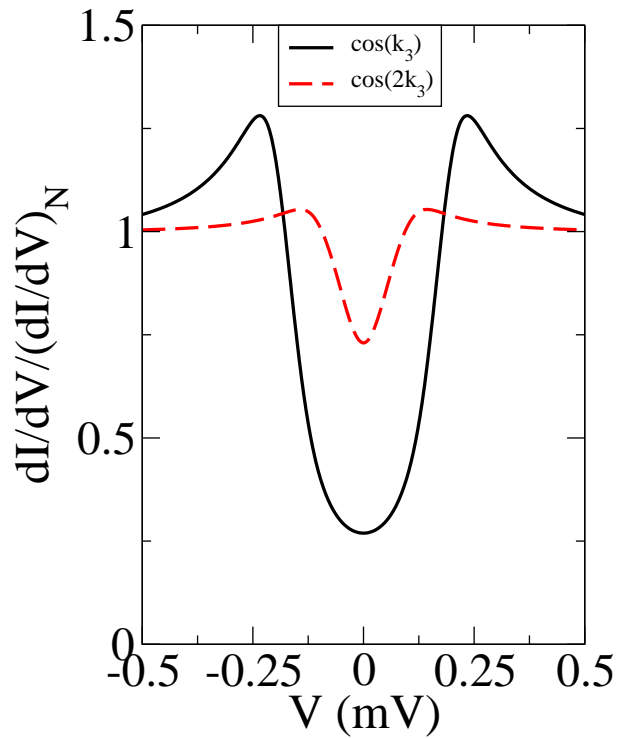}
\caption{Predictions for c-axis NS tunnelling with calcium as the normal metal
are shown.}
\end{figure}
reasonably well-developed coherence peaks, at a magnitude
of approximately $250 \mu V$, and a comparatively low
ZDOS ($\simeq 0.25$) are present.  However, for the $\cos(2ck_{3})$ case,
the coherence peaks are much broader and occur at significantly
lower bias voltages ($V \simeq 150 \mu V$).  In addition, the ZDOS is
much higher - approximately 0.75.  These differences are sufficiently robust
that we believe that a calcium-based tunnel junction experiment, if technically
feasible, would be
sufficient to distinguish between these two order parameters.

For a final experiment of interest, we have computed the T=0 conductance
for the $\cos(k_{3})$ case for three metals commonly used in tunneling
experiments - beryllium, gold, and aluminum - 
and show the results in Figure 9.  As in the previous plot, the coherence
\begin{figure}[h!]
\includegraphics[width=6cm]{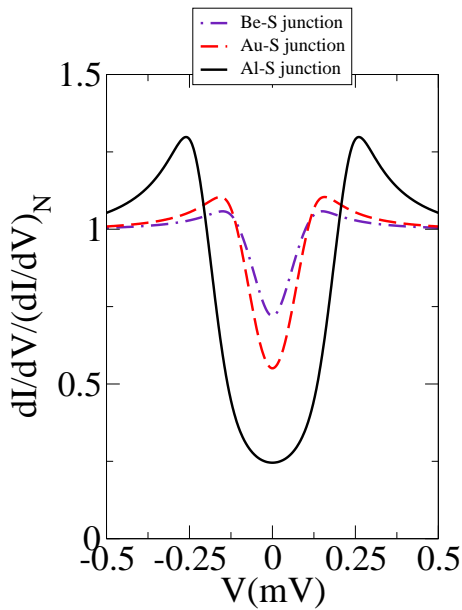}
\caption{Predictions for c-axis NS tunnelling for $\cos(ck_{3})$ with beryllium, gold and aluminum as the normal metal
are shown.}
\end{figure}
peak height and energy vary greatly from one metal to another, with concomitant
variations in the ZDOS.  We believe that such experiments would be a good validation of the basic model employed in this paper, and would provide useful 
information regarding the assumption
that only very few states on the UPd$_2$Al$_{3}$ Fermi surface contribute to
the tunneling conductance.

\section{Conclusion}

We have analyzed the tunnelling spectroscopy of the heavy-fermion
superconductor UPd$_{2}$Al$_{3}$, and studied four possible gap functions for
this material: $\cos(ck_{3}),\,\,\cos(2ck_{3}),\sin(ck_{3})$, and 
$\exp(i\phi)\sin(ck_{3}).$  We find that the last two gap functions would be 
expected to show zero-bias conductance peaks in c-axis tunnelling 
experiments as performed by Jourdan et al \cite{jourdan}; these peaks
were not observed and therefore these order parameters can be excluded.
We further find that the $\cos(ck_{3})$ order parameter 
gives an excellent fit to the experimental data of Jourdan at all temperatures,
with a minimum of fitting parameters and a $\Delta(T)$ appropriate for
the weak-coupling BCS theory of this material.  Finally, we propose a
definitive experiment to distinguish between the $\cos(ck_{3})$ and 
$\cos(2ck_{3})$ order parameters.

\end{document}